\PassOptionsToPackage{table}{xcolor}
\documentclass{fairmeta}

\usepackage{amsfonts}
\usepackage{amssymb}
\usepackage{amsmath}
\usepackage{algorithm}
\usepackage{algpseudocode}
\usepackage{xspace}

\algrenewcommand\algorithmicindent{0.9em}

\usepackage{listings}
\usepackage{inconsolata}

\definecolor{lstbg}{rgb}{0.98,0.98,0.98}
\definecolor{lstframe}{rgb}{0.80,0.80,0.80}
\definecolor{lstkeyword}{rgb}{0.00,0.25,0.62}
\definecolor{lstbuiltin}{rgb}{0.42,0.13,0.55}
\definecolor{lststring}{rgb}{0.60,0.15,0.10}
\definecolor{lstcomment}{rgb}{0.35,0.45,0.35}
\definecolor{lstnumber}{rgb}{0.55,0.55,0.55}
\definecolor{lstdecorator}{rgb}{0.65,0.40,0.00}

\lstdefinestyle{python}{
  language         = Python,
  basicstyle       = \ttfamily\footnotesize,
  keywordstyle     = \color{lstkeyword}\bfseries,
  keywordstyle     = [2]\color{lstbuiltin},
  keywordstyle     = [3]\color{lstdecorator},
  stringstyle      = \color{lststring},
  commentstyle     = \color{lstcomment}\itshape,
  identifierstyle  = \color{black},
  morekeywords     = {async,await,match,case,nonlocal,yield,as,with,assert,
                      lambda,del,pass,raise,from,import,global,True,False,None},
  morekeywords     = [2]{self,cls,super,dataclass,field,range,len,print,
                         enumerate,zip,isinstance,int,str,bool,float,Any,
                         Callable,Optional,List,Dict,Tuple},
  morekeywords     = [3]{@dataclass,@property,@staticmethod,@classmethod,
                         @abstractmethod,@torch,@wraps},
  numbers          = left,
  numberstyle      = \tiny\color{lstnumber},
  numbersep        = 7pt,
  stepnumber       = 1,
  xleftmargin      = 1.6em,
  framexleftmargin = 1.6em,
  frame            = tb,
  framerule        = 0.4pt,
  rulecolor        = \color{lstframe},
  backgroundcolor  = \color{lstbg},
  breaklines       = true,
  breakatwhitespace= true,
  postbreak        = \mbox{\textcolor{lstnumber}{$\hookrightarrow$}\space},
  showstringspaces = false,
  tabsize          = 4,
  columns          = fullflexible,
  keepspaces       = true,
  upquote          = true,
  captionpos       = b,
  abovecaptionskip = 4pt,
  belowcaptionskip = 4pt,
  aboveskip        = 8pt,
  belowskip        = 8pt,
  escapeinside     = {(*@}{@*)},
  literate         =
    {->}{{$\rightarrow$}}2
    {<=}{{$\leq$}}2
    {>=}{{$\geq$}}2
    {!=}{{$\neq$}}2,
}

\usepackage{tikz}
\usepackage{makecell}
\usepackage{threeparttable}
\usepackage{tabularx}
\usepackage[inline]{enumitem}

\newcommand{\fancycircle}[1]{%
  \tikz[baseline=(char.base)]
  \node[
    shape=circle,
    draw=black,
    fill=gray!6,
    text=black,
    inner sep=0pt,
    minimum size=1.05em,
    font=\small
  ] (char) {#1};%
}

\usepackage{comment}
\usepackage{todonotes}

\newcommand{\sys}{WeightBridge\xspace}
\newcommand{\intsys}{internal framework\xspace}

\title{\sys: An Efficient Weight Transfer Library for Reinforcement Learning}

\author[1,2,*]{Xuanlin Jiang}
\author[1]{Samuel Hsia}
\author[1]{Michael Kuchnik}
\author[1]{Zachary DeVito}
\author[2]{Minlan Yu}
\author[1]{Carole-Jean Wu}

\affiliation[1]{FAIR at Meta}
\affiliation[2]{Harvard University}
\contribution[*]{Work carried out at Meta}

\abstract{%
  Weight transfer --- the propagation of updated parameters from trainers to rollout generators --- is becoming an important performance bottleneck in reinforcement learning (RL) systems for LLMs.
The central challenge is supporting the diverse trainer and rollout layouts and synchronization requirements of modern RL workloads without sacrificing efficiency.
Existing solutions are efficient under some configurations but perform poorly or lack support under others. 
We present \sys, a flexible, efficient weight-transfer library designed to deliver high performance across diverse RL configurations.
\sys first automatically extracts the correspondence between trainer and rollout weight layouts, then plans and executes redundancy-free and load-balanced weight transfer.
It exposes a small, general API while coordinating workers across diverse synchronization modes.
Across configurations spanning different models, parallelization layouts, and synchronization modes, \sys reduces average GPU stall time by up to $42\times$ over the state-of-the-art open-source RL framework and achieves high
performance in all settings.
A coding agent was able to integrate \sys into two different RL frameworks without manual guidance, demonstrating the generality and ease of use of its APIs.%
}

\correspondence{Xuanlin Jiang at \email{xjiang@g.harvard.edu}, Minlan Yu at \email{minlanyu@g.harvard.edu}, Carole-Jean Wu at \email{carolejeanwu@meta.com}}

\begin{document}

\maketitle

\section{Introduction} 
\label{sec:intro}
Reinforcement learning (RL) post-training enhances LLMs with capabilities such as mathematical reasoning \citep{deepseekmath, dapo, rlincorrect} and software engineering \citep{rlgpt, swerl} and is becoming an increasingly important workload in modern data centers.
RL alternates between rollout workers that generate trajectories and trainer workers that use those trajectories to update the model.
\emph{Weight transfer} propagates the resulting model weights from the trainers back to the rollout workers.

In RL frameworks, weight transfer has become critical due to two scaling trends.
First, sparse MoE models increase the number of parameters transferred relative to the computation performed per token.
Second, GPU compute throughput is improving faster than inter-node RDMA bandwidth.
These trends shift the RL bottleneck toward communication; in a deployed production workload, we observe GPUs stalled on weight transfer for up to 27\% of total execution time.
This bottleneck prompted recent optimization efforts from Perplexity, ByteDance, Anyscale, and others
\citep{fabriclib,tensorhub,raydirect,primerlwt,awex}.

Achieving efficient weight transfer is non-trivial because modern RL workloads span a rapidly expanding configuration space \citep{verl,areal,miles,asyncflow,laminar,streamrl}.
Training and inference engines are optimized independently and may use different GPU allocations, parallelization strategies, and runtime kernel tensor formats.
As a result, the same logical weight may be partitioned, replicated, reordered, packed, transposed, or quantized differently on the two sides \citep{megatron-lm, alpa, alpaserve}.
RL applications also differ in their tolerance for stale trajectories, requiring different trainer--rollout synchronization modes to balance policy freshness and execution efficiency
\citep{prosperitybeforecollapse,stabilizing-offpolicy,learnhardproblems}.
Therefore, weight transfer must efficiently redistribute and transform weights while coordinating workers across arbitrary combinations of model layouts and synchronization requirements.

Existing weight-transfer mechanisms cover only slices of this configuration space.
TensorHub \citep{tensorhub} reaches near-ideal RDMA throughput only when trainer and rollout layouts are identical and asynchronous execution is acceptable.
Miles \citep{miles} supports synchronous mode RL but loses efficiency as rollout replication grows.
At 16 rollout replicas, its optimized P2P path takes $34\times$ as long as our solution for end-to-end weight transfer.

Supporting diverse configurations at near-hardware performance requires solving the following three challenges.
First, the system must automatically recover the element-wise correspondence between heterogeneous trainer and rollout layouts without model- or backend-specific rules.
Second, it must globally plan the resulting many-to-many transfer so that replicated weights do not cause redundant traffic or load imbalance, and then execute the plan while saturating the underlying network.
Third, it must coordinate workers correctly under different synchronization modes without interfering with the control flows of either the training or inference backend.

We present \sys, a flexible and efficient weight-transfer library that automatically discovers weight correspondence, plans and executes non-redundant transfers, and coordinates workers across different synchronization modes.
\sys is easy to integrate: in an agentic development experiment, a coding agent added it to Miles in under 30 minutes and achieved a $20\times$ weight-transfer speedup.
\sys supports every evaluated Miles configuration, covering models from $16$B to more than $1$T parameters, deployments from $2$ to $64$ H100 nodes, varied trainer/rollout allocations and parallel layouts, and all four synchronization modes.
Across these configurations, \sys reduces average GPU stall time by up to $42\times$ over Miles' optimized native mechanisms.
We also integrate \sys into an internal RL framework with a different backend stack, where it improves end-to-end weight-transfer time by up to $11\times$.

In summary, this paper makes the following contributions:
\begin{itemize}
    \item We design \sys, a generic and efficient weight-transfer library that supports diverse RL configurations while driving performance towards the theoretical lower-bound.

    \item We demonstrate \sys's performance across diverse RL configurations, reducing average GPU stall time by up to $42\times$.

    \item We demonstrate \sys's generality through agentic integration into Miles and deployment in an internal RL framework.
\end{itemize}

The rest of the paper is organized as follows:
\S~\ref{sec:background} describes weight transfer and the diversity of modern RL configurations. \S~\ref{sec:design} introduces the design of \sys.
\S~\ref{sec:eval} evaluates \sys, \S~\ref{sec:discussion} outlines extensions, and finally related work and conclusion are covered in \S~\ref{sec:related} and \S~\ref{sec:conclusion}.

\section{The Weight Transfer Design Space}
\label{sec:background}
This section introduces the \textit{what}, \textit{where}, \textit{when}, and \textit{how} of efficient weight transfer (\S~\ref{sec:wtinrl}), the large and diverse configuration space. (\S~\ref{sec:rl-config}), and \textit{why} it is difficult for existing systems to match all needs (\S~\ref{sec:challenge}).
This design space motivates our system (\S~\ref{sec:design}), which aims to cover the gaps.

\begin{figure}[t]
    \centering
    \includegraphics[trim={0.5cm 0.5cm 0.5cm 0.5cm}, width=0.8\linewidth]{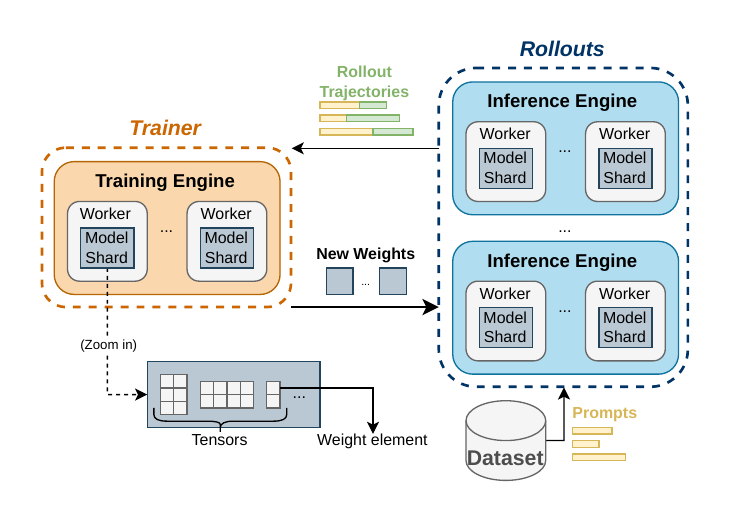}
    \captionsetup{labelfont=bf}
    \caption{\textbf{RL workflow.} Modern RL frameworks typically use only one training engine but multiple inference engines. Training requires strict synchronization between the workers, while inference engines can run independently.}
    \label{fig:rl-workflow}
\end{figure}

\begin{figure*}[t]
  \centering
  \captionsetup{skip=3pt}
  \captionsetup[subfigure]{skip=2pt}
  \begin{minipage}[t]{0.49\textwidth}
    \begin{minipage}[t][4.5cm][t]{\linewidth}
      \centering
      \includegraphics[
      trim={0.5cm, 0.4cm, 0.5cm, 0.4cm}, clip,
      height=4.0cm]{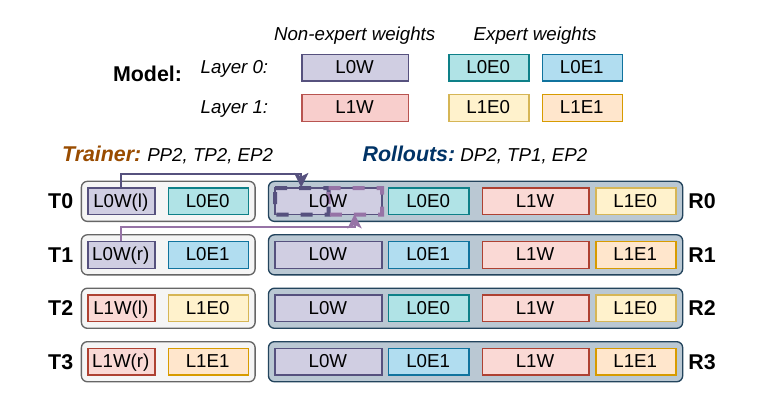}
      \vfill
    \end{minipage}
    \captionof{figure}{\textbf{Complex correspondence between weight elements.}
      T0--T3 are trainer workers, and R0--R3 are rollout workers.
      ``l'' and ``r'' denote parts of the original tensor.}
    \label{fig:example}
  \end{minipage}\hfill
  \begin{minipage}[t]{0.49\textwidth}
    \begin{minipage}[t][4.5cm][t]{\linewidth}
      \centering
      \begin{subfigure}[t]{0.53\linewidth}
        \centering
        \includegraphics[
        trim={0.5cm, 0.4cm, 0.5cm, 0.4cm}, clip,
        height=4.0cm]{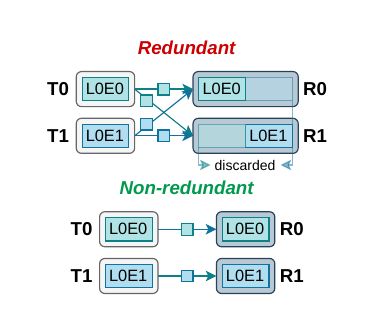}
        \caption{Redundant transfer}
        \label{fig:redundant}
      \end{subfigure}\hfill
      \begin{subfigure}[t]{0.46\linewidth}
        \centering
        \includegraphics[
        trim={0.5cm, 0.5cm, 0.5cm, 0.5cm}, clip,
        height=4.0cm]{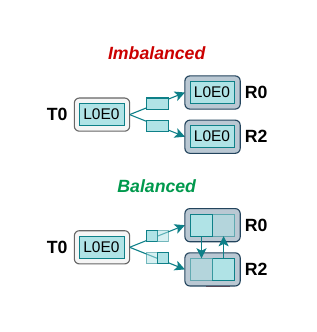}
        \caption{Imbalanced channels}
        \label{fig:imbalanced}
      \end{subfigure}
      \vfill
    \end{minipage}
    \captionof{figure}{\textbf{Two dimensions of efficient weight transfer.}
      Inefficient transfers send redundant weights or poorly balance load
      across channels.}
    \label{fig:wt-ineff-eg}
  \end{minipage}
\end{figure*}

\subsection{Weight transfer in RL}
\label{sec:wtinrl}

A reinforcement learning workflow (Figure~\ref{fig:rl-workflow}) involves rollout workers that generate response trajectories using an LLM inference engine. Trainer workers post-process the trajectories, compute rewards and losses, then update the model weights.

In the RL workflow, weight transfer propagates updated model weights from trainer workers to rollout workers. 
Its performance is becoming increasingly critical.
The increasing sparsity in MoE models makes the volume of transferred weights grow faster than per-token computation, 
and GPU compute throughput improves faster than RDMA bandwidth.
These trends shift the RL bottleneck toward communication, motivating many companies to work on weight-transfer optimizations. 
These companies include Perplexity \citep{fabriclib}, ByteDance \citep{tensorhub}, Anyscale \citep{raydirect}, PrimeIntellect \citep{primerlwt} and Ant Group \citep{awex}.
Weight transfer involves three key steps, corresponding to \textit{what} weights are transferred \textit{where}, \textit{how} the weight transfer happens, and \textit{when} the weights become available:
\begin{enumerate*}[label=\protect\fancycircle{\arabic*}]
    \item\label{itm:layout} weight layout discovery,
    \item\label{itm:transfer} weight transfer planning, and
    \item\label{itm:coordination} worker synchronization and weight transfer scheduling.
\end{enumerate*}
Today's frameworks often make rigid assumptions about one or more of these steps, resulting in lost flexibility or performance.

\ref{itm:layout} \textbf{Weight layout discovery} obtains information about which weight elements each trainer worker holds and which elements each rollout worker requires, and how these elements correspond with each other.
With the layouts of both trainers and rollouts, the system can determine the \textit{correspondence} between them, mapping trainer weights to rollout weights.

\ref{itm:transfer} \textbf{Weight transfer planning} uses layout information (\ref{itm:layout}) to guide how data is routed between workers through the network.
It also decides what data plane operations need to be performed, including RDMA transfers and GPU kernels that transform tensors from training to inference format.

\ref{itm:coordination} \textbf{Worker synchronization and weight transfer scheduling} determines how weight transfer (\ref{itm:transfer}) interacts with the rest of the RL execution loop: 
when computation should pause or continue, how workers should synchronize for communication, 
and how the system should preserve the intended RL semantics across different synchronization modes.

\subsection{Diverse RL configurations}
\label{sec:rl-config}

\noindent\textbf{Diverse model-layout configurations for training and inference.} 
Reinforcement learning frameworks include two key components: 
training backend engines such as Megatron-LM \citep{megatron-lm} and PyTorch FSDP~\citep{torchfsdp}; rollout inference engines such as vLLM~\citep{vllm} and SGLang~\citep{sglang}.
To build an RL framework for a model, developers need to make many model layout decisions.
These decisions include how to allocate GPUs to the trainer and rollout workers, and then the configurations within each training and rollout framework. 
The choice of configurations depends on the models and the hardware.

For training, the model layout configurations include five-dimensional parallelism (data, tensor, pipeline, sequence, and expert parallelisms) \citep{megatron-lm, moe} and state sharding (e.g., ZeRO~\citep{zero}).
The best configuration is jointly determined by model architecture and size, batch and sequence lengths, GPU memory and compute specifications, and interconnect bandwidth and topology \citep{zero, alpa}. 
Model weights need to be sharded and replicated in different ways across GPUs and nodes under different configurations.
Weight transfer needs to determine the data layout correctly (\ref{itm:layout}) and move the weights efficiently (\ref{itm:transfer}), regardless of what the configuration is.

Inference has different objectives and memory profiles, yet its layout configurations (\ref{itm:layout}) are similarly diverse: 
tensor- and pipeline-parallelism degrees, number of replicas, PD-disaggregation \citep{distserve} and CPU offloading plans \citep{flexgen}.
These configurations vary with model size, prompt characteristics, latency targets, and available GPU, CPU, and storage resources \citep{alpaserve}.  

\noindent\textbf{Training and rollout layouts are different within a job.}
Training and rollout engines can adopt different layouts (\ref{itm:layout}) within the same job.
A single training replica often requires more devices than a rollout replica because it must store gradients, activations, and optimizer states in addition to model parameters.
Training can also employ larger pipeline- and expert-parallelism degrees because large training batches better amortize pipeline bubbles and collective-communication overheads, 
whereas rollout engines generally favor configurations optimized for decoding latency and request-level replication.
Training-specific layouts such as FSDP~\citep{torchfsdp} or three-dimensional parallelism and inference-specific layouts such as prefill--decode disaggregation~\citep{distserve} further widen this layout mismatch.
As a result, weight transfer (\ref{itm:transfer}) must redistribute parameters across two groups of GPUs with different sharding and replication strategies, rather than simply copying corresponding shards between GPUs.

Layout mismatches (\ref{itm:layout}) between training and inference backends extend beyond parallelism and sharding strategies, as they may use entirely different shapes, layouts and quantization configurations even for the same tensor.
This is because they use different GPU kernels: training favors high-throughput, large-batch GEMMs, whereas autoregressive inference requires low-latency kernels optimized for small, often memory-bound workloads~\citep{scaleinference, deepspeedinf}.
These specialized kernels can require different physical tensor layouts: QUICK, for example, interleaves quantized weights according to Tensor Core thread-level load patterns~\citep{quick}, while TensorRT-LLM packs and interleaves experts' $W_1$ and $W_3$ weights to fuse SwiGLU into the GEMM epilogue~\citep{moedensegemm}.
Logically equivalent training and inference weights may therefore differ in shape, ordering, fusion, packing, and sharding.
Advanced architectures and techniques---including MoE \citep{moe}, LoRA \citep{lora}, MTP \citep{mtp}, HCA \citep{deepseekv4}, and KDA \citep{kimilinear}---all introduce specialized kernels on both training and inference ends, even deepening the discrepancies.
Bridging different tensor representations requires weight transfer (\ref{itm:transfer}) to perform layout conversion (\ref{itm:layout}) for each logical weight, instead of simple bytewise copying.

\noindent\textbf{Diverse synchronization schemes between trainers and rollouts.}
The synchronization mode (\ref{itm:coordination}) between training and inference may vary as workload changes.
In synchronous mode, rollout workers wait for the latest weights before generating new trajectories, preserving policy freshness at the cost of synchronization stalls.
In asynchronous mode, rollout workers continue generation while weight transfer proceeds in the background, improving resource utilization but introducing staleness.
The choice of synchronization mode depends on the applications and learning algorithms \citep{prosperitybeforecollapse,stabilizing-offpolicy}.

RL applications with alignment objectives---such as chat quality, instruction following, safety, and refusal behavior---often involve short, low-variance rollouts evaluated by fast learned reward models, making the cost of fully synchronous, on-policy execution relatively small.
Reasoning workloads, such as mathematics and competitive programming, produce longer and more variable generations, motivating near-on-policy execution with straggler mitigation or bounded staleness.
Agentic workloads---such as repository-level software engineering, deep research, web browsing, and computer use---have rollout latencies dominated by sandboxes, containers, and external services; their long and highly variable episodes make global synchronization barriers particularly expensive. Moreover, different RL applications may use different model families and sizes with different levels of sensitivity to stale trajectories
\citep{prosperitybeforecollapse}.

The choice of sync and async modes (\ref{itm:coordination}) also depends on the RL algorithm \citep{prosperitybeforecollapse,stabilizing-offpolicy}.
Classic algorithms like PPO \citep{ppo} or GRPO \citep{deepseekmath} require synchronized training and inference execution.
Prior works improve tolerance to staleness through importance sampling, which compensates for differences between the behavior and current policies, and trust-region methods, which limit policy drift.
However, these techniques are not always preferred over classic algorithms, because they incur additional estimation and tuning costs.
RL systems therefore require configurable synchronization modes to satisfy these diverse freshness and efficiency requirements.
Weight transfer must accordingly enforce different rules for computation overlap and worker synchronization across these modes, complicating its coordination with trainer and rollout execution.

In summary, RL workloads use diverse training and inference layouts and synchronization modes.
To support these diverse settings, a booming set of RL frameworks and weight-transfer mechanisms have been proposed to adapt to the increasingly diverse RL configuration space (Table~\ref{tab:weight-transfer-comparison}).

\subsection{Challenges for supporting diverse settings }
\label{sec:challenge}
\begin{table*}[t]
\centering
\captionsetup{position=bottom, labelfont=bf, labelsep=period}
\footnotesize
\setlength{\tabcolsep}{6pt}
\renewcommand{\arraystretch}{1.3}

\begin{tabularx}{\textwidth}{@{}lXlc@{}}
\toprule
\textbf{Method}
& \textbf{Configurations as fast as \sys}
& \textbf{Limitation elsewhere}
& \textbf{Sync. modes} \\
\midrule

\makecell[l]{Miles broadcast~\citep{miles},\\ prime-rl NCCL~\citep{primerl},\\
             VeRL~\citep{verl},\\ AReaL~\citep{areal}}
& \(\mathrm{Trainer}\{\mathrm{DP,PP}\} \rightarrow \mathrm{Rollout}\{\mathrm{DP}\}\)
& redundant transfer & Sync only \\

\addlinespace
Laminar~\citep{laminar}
& \(\mathrm{Trainer}\{\mathrm{DP,PP,EP}\} \rightarrow \mathrm{Rollout}\{\mathrm{DP}\}\)
& redundant transfer & Async only \\

\addlinespace
\makecell[l]{Miles P2P~\citep{milesp2p},\\ prime-rl P2P~\citep{primerlwt},\\
             RDT~\citep{raydirect}}
& \makecell[l]{\(\mathrm{Trainer}\{\mathrm{DP,PP,EP}\} \rightarrow\)\\
             \(\mathrm{Rollout}\{\mathrm{DP\ (low\ degree)},\mathrm{EP}\}\)}
& \makecell[l]{
single-source bottleneck \\ as rollout DP grows
} & Sync only \\

\addlinespace
fabric-lib~\citep{fabriclib}
& \makecell[l]{\(\mathrm{Trainer}\{\mathrm{DP,PP,EP}\} \rightarrow\)\\
             \(\mathrm{Rollout}\{\mathrm{DP\ (low\ degree)},\mathrm{EP}\}\)}
& \makecell[l]{
single-source bottleneck \\ as rollout DP grows
} & Async only \\

\addlinespace
TensorHub~\citep{tensorhub}
& Identical trainer--rollout layouts
& collocated layout conversion & Async only \\

\addlinespace
\rowcolor{blue!5}
\sys
& \makecell[l]{\(\mathrm{Trainer}\{\mathrm{DP,PP,EP}\} \rightarrow\)\\
             \(\mathrm{Rollout}\{\mathrm{DP},\mathrm{EP}\}\)}
& \textemdash & Sync and async \\

\bottomrule
\end{tabularx}

\caption{
Comparison of weight-transfer mechanisms (\ref{itm:transfer}) across parallelization configurations (\ref{itm:layout}) and synchronization modes (\ref{itm:coordination}).
Each library is efficient only in part of the space, whereas \colorbox{blue!5}{\sys} is efficient throughout.
We assume an MoE model with disaggregated trainer and rollouts; DP/PP/EP describe inter-node layout.
}
\label{tab:weight-transfer-comparison}
\end{table*}
Supporting diverse RL configurations requires more than fast data movement. 
As Table~\ref{tab:weight-transfer-comparison} shows, existing mechanisms are typically efficient only in a subset of the layout and synchronization configurations. 
These limitations expose three challenges. 
\ref{itm:layout} To eliminate transfer redundancy for diverse model layouts, weight transfer must discover element-wise correspondence across heterogeneous trainer and rollout layouts without relying on any model- or backend-engine-specific assumptions.
\ref{itm:transfer} To ensure load balance for diverse replication patterns and heterogeneous links, it must globally plan the resulting many-to-many transfers to eliminate redundant communication and balance load across replicated workers. 
\ref{itm:coordination} To support diverse synchronization modes and backend engines, it must provide a set of APIs with minimal coupling to training and inference engines.
We discuss these challenges in turn.

\noindent\ref{itm:layout}~\textbf{Weight layout discovery: deriving element-wise correspondence across heterogeneous layouts.}
Weight transfer is sending weight shards from corresponding trainers to the rollouts that need them.
Doing so requires knowing the correspondence between weight elements across the workers before transmitting the data.
However, this is not a trivial task, 
because different layout configurations of the backend engines could result in entirely different layouts, and thus create a correspondence problem between trainers and rollouts.
For example, in Figure~\ref{fig:example}, if the trainer and rollout both use (TP2, EP2, PP2), their layouts are identical and the correspondence between them is simply one-to-one parallel across the workers.
However, when rollouts are changed to (TP1, EP2, DP2), the correspondence becomes complex, and every rollout worker requires some weight from every trainer worker, causing a cross-rank correspondence.
As the number of tensors increases and different parallelization configurations are mixed, the correspondence graph becomes more complex.

If we do not know the exact correspondence before transferring the weights, every rollout worker should receive a full copy of the model for correctness, as used in many existing frameworks \citep{miles,verl,areal,laminar}.
As Figure~\ref{fig:redundant} shows, this leads to redundant transfer and hurts efficiency.
To avoid some redundancy, other solutions discover correspondence based on specific models and backend engines.
Awex \citep{awex} finds the correspondence through a hard-coded look-up table, which records a set of rules describing how the correspondence between each model and each pair of backends should be calculated.
This approach cannot support any model or backend that is not recorded in the lookup table, and will fail silently when an update to an existing backend changes the way it loads some models.
Prime-RL \citep{primerlwt} instead extracts the correspondence by recording the PyTorch operations applied when converting trainer weights to checkpoint format and converting checkpoint-format weights to rollout weights.
Although this approach no longer relies on specific models, it will fail if any non-PyTorch operation is applied in the pipeline.

Therefore, our goal is to develop a generic layout-discovery mechanism that automatically derives the weight element correspondence across diverse layout configurations without relying on any model- or backend-engine-specific assumptions.
This correspondence information will provide guidance for transfer planning, avoiding redundant transfer.

\noindent\ref{itm:transfer}~\textbf{Weight transfer planning: global routing across many-to-many correspondence.} 
Another challenge for weight transfer is that weight tensors are often replicated across workers. Thus weight transfer is no longer point-to-point between two workers, but a collective communication between two replica groups.
Different tensors may involve entirely different groups, making the routing problem even more complex.
As shown in Figure~\ref{fig:imbalanced}, if both R0 and R2 pull L0E0 only from T0, there is a load imbalance:
The egress volume from the sender to different receivers accumulates, 
while the channels between the receivers remain idle, leading to a single-source bottleneck.
Many existing systems \citep{milesp2p,raydirect,primerlwt,fabriclib,awex} exhibit this bottleneck: as Table~\ref{tab:weight-transfer-comparison} shows, their performance worsens as RDP increases, because they make rollout workers individually decide on where to pull the weights without any global coordination.
Efficient weight transfer thus requires global planning based on the correspondence between weight elements.
It needs to eliminate redundant transfer while achieving communication load balance across trainer and rollout workers,
where communication load for each worker needs to account for aggregate contribution from each replica group it is in.
These objectives become harder to satisfy when workers are connected through heterogeneous links.

\noindent\ref{itm:coordination}~\textbf{Worker synchronization and weight transfer scheduling: supporting diverse synchronization modes with minimal coupling.}
The APIs and control plane workflow of weight transfer must be carefully designed to support the diversity of RL configurations and backends.
Two basic requirements need to be satisfied.
First, the APIs must generally support different synchronization modes for diverse RL workloads.
In synchronous mode, they must ensure that all trainer and rollout workers enter weight transfer simultaneously and return only after it fully completes.
In asynchronous mode, the trainer-side API should allow training to resume while transfer continues in the background, overlapping computation and communication.
Similarly, rollout workers should continue generation until the new weights have arrived in local buffers and can be loaded directly.
Existing systems commonly target only one mode: Miles \citep{miles} uses send--receive-style APIs with blocking semantics, whereas TensorHub \citep{tensorhub} uses load--store-style APIs designed for asynchronous execution.

Second, to support different backend engines, the weight-transfer control flow should not rely on specific control-planes features of the backends and should have an independent control plane instead.
For example, all workers belonging to the same inference engine must apply a weight update at the same inference step to avoid racing or deadlock.
Existing frameworks \citep{miles,verl} commonly achieve this by piggybacking weight-transfer signals on the inference backend's control plane, propagating them through the same channels as other control messages, so that workers receive them.
This tight coupling complicates integration with other backends, which may use different control-plane architectures and protocols.
The same principle also applies to training backends.
Therefore, a general weight-transfer design should avoid backend-specific control-plane dependencies and require minimal modifications to training and inference engines.

\section{\sys}
\label{sec:design}

\begin{figure}[t]
    \centering
    \includegraphics[trim={0.7cm 0.8cm 0.7cm 0.7cm}, clip, width=0.7\linewidth]{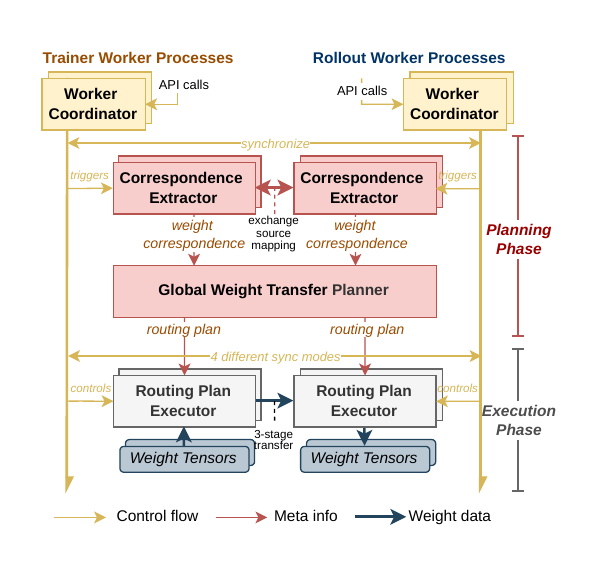}
    \captionsetup{labelfont=bf}
    \caption{\textbf{The \sys workflow}, which extracts weight layout correspondence between trainer and rollout workers (\ref{itm:layout}), plans the transfer (\ref{itm:transfer}), and executes it using application synchronization semantics (\ref{itm:coordination}).}
    \label{fig:design}
\end{figure}

Motivated by the gaps in existing systems due to the large design space of weight transfer (\S~\ref{sec:background}), we introduce \sys, which aims to support all weight transfer needs.
As shown in Figure~\ref{fig:design}, \sys is organized around four components: the \emph{automatic correspondence extractor}, \emph{global weight transfer planner}, \emph{routing plan executor}, and \emph{worker coordinator}.
The automatic correspondence extractor determines the weight layout on each worker and derives the element-wise correspondence between trainer and rollout workers (\ref{itm:layout}).
The global weight transfer planner then converts this correspondence into an optimized routing plan (\ref{itm:transfer}).
Whenever a new model version is produced, the worker coordinator triggers the execution phase (\ref{itm:coordination}), in which the routing plan executor follows the precomputed plan to transfer the updated weights.

\S~\ref{sec:correspondence} describes the automatic correspondence extractor.
By analyzing each backend's weight-loading behavior, it derives element-wise correspondence across heterogeneous layouts without model- or backend-specific information.

\S~\ref{sec:transfer} presents the global weight transfer planner and routing plan executor.
The planner constructs a topology-aware, non-redundant routing plan that balances communication across replicated workers.
The executor reuses this plan for every model update and applies pipelining and batching optimizations to maximize bandwidth utilization.
Together, they drive end-to-end transfer time toward the theoretical lower bound.

Finally, \S~\ref{sec:api} presents the worker coordinator.
It provides a small, unified interface, supports four synchronization modes, and coordinates workers independently of the backend engines' control planes.

\subsection{Layout Correspondence Extraction}
\label{sec:correspondence}

\begin{figure*}[t]
  \centering
  \captionsetup{skip=3pt}

  \begin{minipage}[t]{0.49\textwidth}
    \begin{minipage}[t][4.85cm][t]{\linewidth}
      \centering
      \includegraphics[trim={0.5cm, 0.5cm, 0.5cm, 0.5cm}, clip, 
      height=4.6cm]{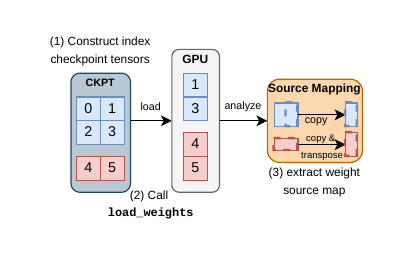}
      \vfill
    \end{minipage}
    \captionof{figure}{\textbf{Outline of source-map extraction.}
      The algorithm runs on every worker; the results are exchanged to
      compute the final weight correspondence.}
    \label{fig:specgen}
  \end{minipage}\hfill
  \begin{minipage}[t]{0.49\textwidth}
    \begin{minipage}[t][4.85cm][t]{\linewidth}
      \centering
      \includegraphics[trim={0.5cm, 0.5cm, 0.5cm, 0.2cm}, clip, 
      height=4.6cm]{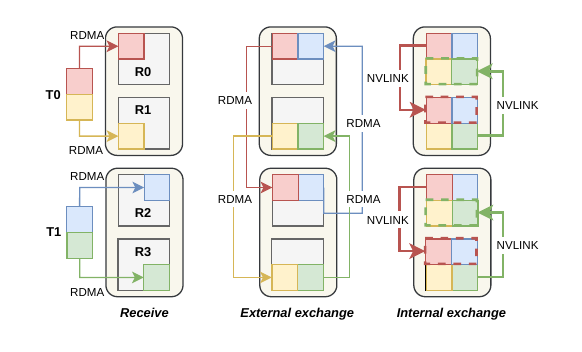}
      \vfill
    \end{minipage}
    \captionof{figure}{\textbf{Example of three-stage transfer.}
      The trainer uses TP2 and the rollouts use DP4.
      T0--T1 are trainer workers; R0--R3 are rollout workers.
      The two exchange stages reduce each rollout node's total ingress
      from six small squares under direct all-to-all exchange to four.}
    \label{fig:3stage}
  \end{minipage}
\end{figure*}

To enable non-redundant weight transfer, \sys must recover the element-wise correspondence between trainer and rollout workers (\ref{itm:layout}).
Existing approaches derive this correspondence from hard-coded model rules or by relying on backend-specific implementations \citep{awex,primerlwt}.
We instead design an \emph{automatic layout correspondence extractor} that treats each worker's \texttt{load\_weights} function as a black box and derives the correspondence without model- or backend-specific information.
The extractor does not even require the workers' parallelization configurations or ranks within their backend engines.

Two key observations motivate our design:
First, the checkpoint of the full model can be used as a bridge between the in-GPU parameters of trainer and rollout workers. 
If we know the weight source map of each worker, that is, which elements it holds in the checkpoint, 
we can compute the overlapping portion of the source maps between each pair of workers to obtain the correspondence. 
Second, the \texttt{load\_weights} function of training and inference backends --- which loads the checkpoint into the GPU parameter tensors during engine startup --- exactly encapsulates this source map information.

Therefore, our solution approaches the problem in two steps.
First, for each trainer and rollout worker, we probe the \texttt{load\_weights} function to obtain its weight source map.
As shown in an example in Figure~\ref{fig:specgen}, we can assign a unique index to each element in the checkpoint.
Calling \texttt{load\_weights} on it will then mark every element in the GPU parameters with its source, thereby revealing the weight source map.
Second, we all-gather the weight source map across all workers, then compute the overlapping portion between each pair of workers to obtain the correspondence of weights.
This approach supports diverse RL model layouts using only each backend's \texttt{load\_weights} function, requiring only that it preserve same-dtype values.
We represent the source map as a sequence of slice-copy operations performed by \texttt{load\_weights}, each optionally involving transpose.

In production, large MoE models~\citep{deepseekv4,kimik3} make the indices costly to store.
For example, representing source indices as 64-bit integers requires at least $4\times$ the memory occupied by FP16 or lower-precision weights, risking running out of GPU memory.
We address this problem by computing the second-order differences (\texttt{diff2}) of the index tensors.
With a careful construction, these differences contain at most 16 nonzero entries per slice-copy operation performed by \texttt{load\_weights}.
In addition, the nonzero entries directly delineate boundaries of these operations,
so an analysis of them is sufficient for recovering the source map.
Splitting indices across multiple loading passes accommodates smaller parameter dtypes without materializing full 64-bit index tensors.
Appendix~\ref{app:loadspec} provides more details.
As a result, \sys is able to automatically find the layout correspondence.

\subsection{Globally Optimized Weight Transfer}
\label{sec:transfer}

Once the element-wise weight correspondence is known, \sys's \emph{global weight transfer planner} determines where each weight element should be sent (\ref{itm:transfer}).
It constructs a topology-aware, non-redundant three-stage routing plan that globally balances traffic across replicated workers.
The \emph{routing plan executor} reuses this plan in every RL iteration, applying a round-based, batched pipeline that saturates RDMA bandwidth and drivew end-to-end transfer toward the theoretical optimum.

\textbf{Generating an efficient routing plan.} 
Given the element-wise weight correspondence, we can optimize weight transfer globally.
As discussed in \S~\ref{sec:challenge}, each weight element may be held by multiple trainer workers and needed by multiple rollout workers.
Moreover, the workers are connected by a hierarchical network: GPUs within a node communicate over high-bandwidth NVLink, whereas cross-node traffic goes through slower RDMA links.
The planner must exploit both replication and network topology to deliver all required weights without sender bottlenecks or unnecessary cross-node traffic.

Two principles guide our design.
First, the trainer workers should collectively transmit only one copy of the model, preventing redundant transfer or sender bottlenecks.
Second, each rollout node should receive over RDMA exactly one copy of each weight element it requires, and no elements that it doesn't need;
additional copies needed by workers within the node should be distributed locally over NVLink.

Following these principles, our planning algorithm constructs a three-stage routing plan.
It first logically de-duplicates tensors on both sides.
If a tensor is replicated on the trainer side, the planner assigns each replica a distinct slice, so the trainer workers collectively send only one copy of the tensor.
For a tensor replicated across rollout workers, it treats each replica as initially needing only a slice to avoid requesting duplicated data from the trainer --- the remaining required data are later exchanged between them.
This de-duplicated view reduces the many-to-many weight correspondence to a one-to-one mapping.
The resulting plan can therefore proceed in three stages:
a \emph{receive} stage routes the assigned slices from trainer workers to rollout workers; an \emph{external-exchange} stage uses RDMA to ensure that each rollout node obtains one copy of every weight element it requires; and an \emph{internal-exchange} stage distributes those weights among workers within each node over NVLink.

Figure~\ref{fig:3stage} illustrates this plan.
Although each rollout worker in the example requires the entire model, each receives only a distinct quarter from the trainers during the receive stage.
During the external exchange, rollout workers exchange their received quarters with their counterparts on the other node, giving each node one copy of all required weights.
Finally, the workers exchange weights locally over NVLink so that each obtains the complete model.

The traffic volumes produced by this routing plan match those required by the following lower bound on end-to-end transfer time:
\begin{equation}
\label{eq:lower-bound}
T_{\min} =
\max\left(
\left\{\frac{M}{B_T}\right\}
\cup
\left\{\frac{M_v}{B_v} : v\in R\right\}
\right),
\end{equation}
where $M$ is the model size, $B_T$ is aggregate trainer egress bandwidth, and $M_v$ and $B_v$ are the deduplicated weight volume and ingress bandwidth of rollout node $v\in R$.
The maximum captures both trainer-egress and rollout-ingress bounds.
Our plan transmits exactly $M$ bytes from trainers and $M_v$ bytes into each rollout node, making it traffic-optimal and able to approach $T_{\min}$ when transfers saturate available bandwidth.

\noindent\textbf{Optimizing the transfer stack.}
A traffic-optimal routing plan can approach $T_{\min}$ only if the data-plane implementation keeps the network links saturated.
Doing so presents two challenges.
First, GPU memory limits the size of the transfer buffers.
In our evaluation setup, a 1T-parameter model occupies more than 60\,GB per GPU, leaving less than 5\,GB for RDMA buffers, so buffering an entire model shard is infeasible.
Second, different tensors from the same trainer may have different destinations and require different layout transformations.
Existing systems handle each tensor with a separate RDMA operation and GPU transformation kernel.
However, modern models can contain hundreds of thousands of tensors \citep{kimik3}, and each worker may need to process thousands to tens of thousands of them.
The resulting per-tensor overhead is substantial: our Miles benchmark \citep{miles} shows that format conversion alone takes 590\,ms, exceeding $3T_{\min}$.

We address both challenges with a round-based, batched transfer pipeline.
The planner partitions the model into approximately equal-sized chunks and transfers one chunk per round, overlapping consecutive rounds of the three-stage transfer to saturate RDMA bandwidth and hide the faster NVLink communication.
Because the complete routing plan is known in advance, we also batch communication and transformation across tensors.
On the trainer side, a packing kernel transforms the tensors planned for the current round into the checkpoint format and coalesces them into a contiguous buffer for each receiver.
On the rollout side, an assembly kernel gathers shards from different senders, transforms them into the required inference format, and writes them into the corresponding model parameters.
Consequently, each worker launches only one or two GPU kernels per round,
In our benchmarks, the packing and assembly kernels account for less than $3\%$ of the end-to-end transfer time and are fully overlapped with RDMA communication, helping drive the overall transfer time toward $T_{\min}$.

\subsection{APIs and Worker Coordination}
\label{sec:api}

\begin{figure*}[t]
  \centering
  \captionsetup{skip=3pt}

  \begin{minipage}[t]{0.49\textwidth}
    \begin{minipage}[t][4.6cm][t]{\linewidth}
\begin{lstlisting}[
  language=Python,
  basicstyle=\ttfamily\scriptsize,
  aboveskip=0pt,
  belowskip=0pt
]
from weightbridge import SenderAdapter, ReceiverAdapter
def trainer_worker_process():
    adapter = SenderAdapter(comm_handles, load_weights, sender_staging=...)
    adapter.connect()
    for i in range(num_iters):
        train_one_step()
        adapter.send_weights()
def rollout_worker_process():
    adapter = ReceiverAdapter(comm_handles, load_weights, receiver_staging=...)
    while True:
        generate_one_step()
        if adapter.poll_requests():
            clean_up_after_weight_update()
\end{lstlisting}
      \vfill
    \end{minipage}
    \captionof{figure}{\textbf{\sys API usage.}
      \texttt{comm\_handles} encapsulates socket information for
      cross-adapter communication.
      \texttt{load\_weights} is probed to generate weight source maps.}
    \label{fig:apicode}
  \end{minipage}\hfill
  \begin{minipage}[t]{0.49\textwidth}
    \begin{minipage}[t][4.6cm][t]{\linewidth}
      \centering
      \vfill
      \includegraphics[
      trim={0.5cm, 0.0cm, 0.5cm, 0.3cm}, clip,
      width=\linewidth]{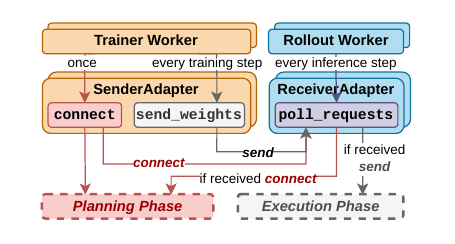}
      \vfill
    \end{minipage}
    \captionof{figure}{\textbf{\sys control plane.}}
    \label{fig:apis}
  \end{minipage}
\end{figure*}

Weight transfer requires coordinated participation from all trainer and rollout workers, with synchronization semantics that vary across RL workloads (\ref{itm:coordination}).
\sys addresses both requirements with a \emph{worker coordinator}.
It exposes a small, unified API for weight transfer planning and execution, supports four synchronization modes, and coordinates workers independently of the backend engines' control planes.

\noindent\textbf{Minimal APIs and configurable synchronization.}

As illustrated in Figure~\ref{fig:apicode}, the worker coordinator exposes three runtime APIs through the \texttt{SenderAdapter} and \texttt{ReceiverAdapter} classes. 
On the trainer side, \texttt{connect} performs one-time correspondence extraction and transfer planning, while \texttt{send\_weights} executes the resulting routing plan for each new model version. 
On the rollout side, \texttt{poll\_requests} polls messages from the trainer and reacts accordingly. It can invoke an optional \texttt{before\_update\_hook} before installing new weights. 
Integration requires only adding these calls at existing framework boundaries: each worker creates an adapter with its communication handles and \texttt{load\_weights} function, trainers call \texttt{connect} at startup and \texttt{send\_weights} after each RL iteration, and rollout workers call \texttt{poll\_requests} between inference steps. 
The same integration points apply across different training and inference backends.

Users select the desired synchronization mode through two adapter options: sender staging (\texttt{SS}) and receiver staging (\texttt{RS}).
For fully synchronous execution, users disable both options; weights are transferred directly between trainer and rollout GPUs, and computation on both sides blocks until the update is installed.
For fully asynchronous execution, users enable both options: \texttt{SS} allows training to resume after the updated weights are staged in host memory, while \texttt{RS} allows inference to continue as incoming weights are staged on the rollout side.
This overlaps transfer with computation on both sides at the cost of potentially greater trajectory staleness.
Users may also enable only \texttt{SS} or \texttt{RS} to overlap transfer with computation on one side, which is useful when training and inference times are imbalanced.
All four modes share the same APIs, routing plan, and data-plane optimizations.

\noindent\textbf{Backend-independent worker coordination.}
Figure~\ref{fig:apis} shows how the worker coordinator synchronizes workers behind the APIs.
All trainer workers invoke \texttt{connect} and \texttt{send\_weights} collectively to trigger the planning and execution phases, respectively.
For either operation, the rank-0 \texttt{SenderAdapter} sends a request to the rank-0 \texttt{ReceiverAdapter} of each rollout engine.
Each rank-0 receiver then relays the request to the other receivers in its engine, ensuring that all workers in the same rollout engine process the request at the same inference step.
Each rollout worker receives and services the request through \texttt{poll\_requests}.
Because these control messages are handled entirely by the worker coordinator, \sys does not depend on or modify the control-plane protocols of the training and inference engines.

\section{Evaluation}
\label{sec:eval}

In this section, we evaluate \sys. We ask two questions: \textbf{(1)} Does \sys maintain high performance across different model scales, parallelization configurations and synchronization modes? \textbf{(2)} Are its APIs general and easy to integrate for different backends? We answer the first question through comprehensive measurements across all these axes, and answer the second through an agentic integration experiment and by reporting our experience integrating \sys with different RL frameworks.
\subsection{Experiment Setup}
\label{sec:eval-setup}

\noindent\textbf{Testbed.}
We run our experiments on an H100 cluster.
Each node has eight NVIDIA H100 80GB HBM3 GPUs connected by NVLink/NVSwitch and 32 100-Gbit/s RDMA-capable NICs, providing 400GB/s of nominal aggregate inter-node bandwidth.

\noindent\textbf{Metrics.}
We primarily report Average GPU Stall Time (AGST) per weight transfer, defined as total GPU stall time divided by the number of GPUs in the cluster. If the stall time is reduced by $\Delta T$ equally across all GPUs, an $n$-iteration RL job will finishes $n\Delta T$ earlier, and throughput increases by $\frac{T_S}{T_S-\Delta T}$, where $T_S$ is the average total training step time before the optimization.

We also measure the End-to-end Weight Transfer Time (EWTT) --- the span from when the first GPU stops computation to when the last GPU resumes.
The ideal EWTT is computed according to Equation~\eqref{eq:lower-bound}.

\noindent\textbf{Baselines.}
We integrate \sys Miles~\citep{miles}, a state-of-the-art open-source RL framework, and compare it with Miles' two native weight transfer mechanisms.

\textit{Broadcast} is Miles' default NCCL broadcast path: trainer workers first all-gather weights to one head rank, then it broadcasts the gathered data to every rollout worker.

\textit{P2P} \citep{milesp2p} is an optimized weight transfer mechanism from Miles: it directly RDMA-writes rollout worker GPUs from the trainer CPU. This eliminates redundant received data, but potentially exposes a single-source bottleneck.

\noindent\textbf{Models.}
We evaluate \sys on representative models across a large range of sizes, including Moonlight-16B-A3B \citep{moonlight}, Qwen3-30B-A3B, Qwen3-235B-A22B \citep{qwen3}, and Kimi-K2-Instruct mixture-of-experts \citep{kimik2} models.
All four checkpoints store their model tensors in BF16.
For the Kimi-K2 workload, we use BF16 instead of the default FP8 configuration parameters for rollout engines to align the precision with the trainer engine.
Misaligned precision can be handled by trainer-side round-by-round gather and quantization in theory, but is currently out of \sys's scope (\S~\ref{sec:discussion}).

\noindent\textbf{Workloads.}
We run the Miles disaggregated RL pipeline with Megatron-LM trainers and SGLang rollout engines.
All four canonical deployments draw prompts from the DAPO-Math dataset.
Moonlight-16B-A3B uses GRPO, while the other three deployments use GSPO. 
The choice of dataset and algorithm doesn't directly affect the metrics we measure.

The scale study uses the four canonical Miles deployments in Table~\ref{tab:real-workload-configurations}; the sensitivity studies use Qwen3-30B-A3B with synthetic placements.

\subsection{Performance Overview for Real Reinforcement Learning Workloads}
\label{sec:real-workload}

\begin{table*}[t]
  \centering
  \setlength{\tabcolsep}{3pt}
  \resizebox{\textwidth}{!}{
  \begin{tabular}{@{}clrrlll@{}}
    \hline
    \textbf{Panel} & \textbf{Model} & \textbf{Parameters} & \textbf{Model size} & \textbf{Cluster} & \textbf{Trainer layout} & \textbf{Rollout layout} \\
    \hline
    (a) & Moonlight-16B-A3B & 16.0 B & 31.9 GB & 2 nodes & 1 node; TP2, EP8 & 1-node engine; TP8, EP8 \\
    (b) & Qwen3-30B-A3B & 30.5 B & 61.1 GB & 4 nodes & 2 nodes; TP4, EP8 & $2\times$1-node engines; TP8, EP8 \\
    (c) & Qwen3-235B-A22B & 235.1 B & 470.2 GB & 16 nodes & 8 nodes; TP4, PP4, CP2, EP16 & $2\times$4-node engines; TP32, EP32 \\
    (d) & Kimi-K2-Instruct & 1.03 T & 2.05 TB & 64 nodes & 32 nodes; TP8, PP8, CP4, EP32 & $8\times$4-node engines; EP32 \\
    \hline
  \end{tabular}}
  \captionsetup{labelfont=bf}
  \caption{\textbf{Real-workload configurations.}
  TP, PP, CP, and EP denote tensor, pipeline, context, and expert parallelism, respectively.
  }
  \label{tab:real-workload-configurations}
\end{table*}

\begin{figure}[t]
  \centering
  \includegraphics[width=0.7\columnwidth]{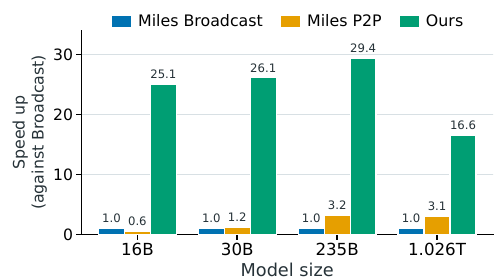}
  \captionsetup{labelfont=bf}
  \caption{\textbf{Speedup over Broadcast across real workloads.}
  Speedup is calculated by dividing the AGST of Broadcast by the corresponding AGST. 
  AGST and EWTT are the same for the baselines because Miles doesn't overlap weight transfer with compute.
  }
  \label{fig:real-workload}
\end{figure}

We evaluate \sys on the four canonical RL deployments summarized in Table~\ref{tab:real-workload-configurations}, scaling from 16B parameters on two nodes to 1T parameters on 64 nodes.
Figure~\ref{fig:real-workload} shows that \sys lowers AGST $25.1\times$, $26.1\times$, $29.4\times$, and $16.6\times$ compared to Miles broadcast and $42.0\times$, $21.0\times$, $9.1\times$, and $5.4\times$ compared to Miles P2P for the 4 deployments, respectively.
Miles P2P pays for trainer-side gathering and CPU staging before a single-source fan-out, while Miles broadcast transfers replica-redundant data and synchronizes a global collective.
Both also expose per-tensor and memory-registration overheads on the critical path.
\sys instead routes each shard directly to a useful destination, exchanges between rollout workers, and pipelines transformation, RDMA, and replica reconstruction.

\sys remains comparable to ideal EWTT across different scales.
For the 16B, 30B, 235B, and 1T deployments, \sys's AGST is $0.90\times$, $0.89\times$, $1.46\times$, and $2.02\times$ ideal EWTT, while its EWTT is $1.27\times$, $1.18\times$, $1.69\times$, and $2.12\times$ ideal, respectively.
Its AGST gets even lower than ideal EWTT at two and four nodes because of communication and computation overlap --- the trainer can exit after handing off its final work, while a rollout engine can enter its stall window only after some incoming work has already progressed.
This effect fades as the routing plan uses more rounds --- at 16 nodes, AGST rises to $1.46\times$ ideal EWTT.
The gap grows at larger scales because additional transfer rounds and rollout engines increase pipeline and control-plane overhead.

\subsection{Varying Model Layouts}
\label{sec:placement}

\begin{figure*}[t]
  \centering
  \captionsetup{skip=3pt}

  \begin{minipage}[t]{0.49\textwidth}
    \begin{minipage}[t][4.75cm][t]{\linewidth}
      \centering
      \includegraphics[height=4.5cm]{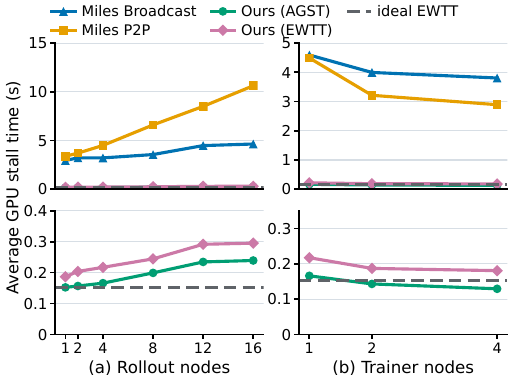}
      \vfill
    \end{minipage}
    \captionof{figure}{\textbf{Varying trainer/rollout nodes.}
      (a) Scaling rollout nodes with one trainer node fixed.
      (b) Scaling trainer nodes with four rollout nodes fixed.
      Each node holds a full model replica with intra-node layout TP8EP8.
      The lower plots zoom in on the y-axis.
      Original Broadcast degrades sharply beyond eight rollout nodes because
      it launches too many concurrent NCCL broadcasts; in (a), we use a
      patched version that batches these operations.}
    \label{fig:sensitivity-allocation}
    \label{fig:vary-rollout}
    \label{fig:vary-trainer}
  \end{minipage}\hfill
  \begin{minipage}[t]{0.49\textwidth}
    \begin{minipage}[t][4.75cm][t]{\linewidth}
      \centering
      \includegraphics[height=4.5cm]{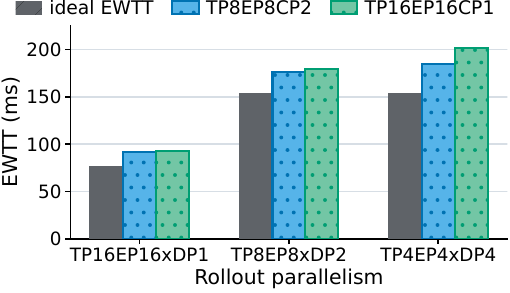}
      \vfill
    \end{minipage}
    \captionof{figure}{\textbf{Varying parallelism.}
      The legend gives trainer parallelism, and the x-axis gives rollout
      parallelism.
      \sys's EWTT tracks ideal EWTT across all six combinations.}
    \label{fig:sensitivity-dp}
  \end{minipage}
\end{figure*}

\noindent\textbf{Scalability with rollout replicas.} Figure~\ref{fig:vary-rollout} compares the scalability of \sys against the baselines.
We fix the number of trainer replicas to one and increase the number of rollout nodes, each node holding a replica of the model.
Miles P2P's AGST grows almost linearly due to its single-source bottleneck, whereas Miles Broadcast remains relatively stable since NCCL broadcast relays between recipients to balance the traffic. 
\sys exhibits sublinear growth in AGST, mainly driven by larger pipeline bubbles and the quadratically growing number of control-plane messages as the number of rollout nodes increases.

\noindent\textbf{Impact of trainer replicas.} We fix rollout nodes to 4 and vary the number of trainer nodes instead. As shown in Figure~\ref{fig:vary-trainer}, both Broadcast and \sys are relatively stable because their weight transfers are both receiver-ingress bounded.
However, for P2P, the total egress for the trainer is $4\times$ the model size, making the transfer time sender-egress bounded, so AGST drastically increases when trainer node count is reduced.

\noindent\textbf{Impact of parallelism degrees.} 
We fix the node allocation for the 30B model deployment and vary DP degrees for both trainer and rollouts, decreasing TP and EP accordingly.
As shown in Figure~\ref{fig:sensitivity-dp}, \sys's measured EWTT tracks the ideal EWTT well across all configurations.
Trainer parallelism has little impact because weight transfer is rollout-ingress-bound.
Increasing rollout DP (RDP) from 1 to 2 doubles per-rollout-node weight demand from $0.5M$ to $M$, while RDP 2 to 4 only adds a second replica per node without changing $M_v$.
Ideal EWTT follows, doubling then flattening.

\subsection{API Evaluation}

\noindent\textbf{Integration to an existing RL framework.}
We quantify the integration effort of \sys through an agentic development experiment. 
We also ask whether a coding agent can come up with optimizations comparable to \sys independently.
We run two isolated Claude Opus~5 agents, each with a fresh agent home and an isolated workspace.
The experiment uses the basic 4-node (30B) setup from Figure~\ref{fig:real-workload}.
The agents receive the same Miles snapshot and are asked to optimize weight transfer for 2 workloads --- MoE Qwen3-30B-A3B uses the same layout as in \S~\ref{sec:real-workload}, while dense Qwen3-32B uses a TP8/PP2 trainer and two TP8 rollout engines.

Agent~A additionally receives \sys as a read-only library and is instructed to integrate it with Miles without further performance tuning.
\sys's documentation only contains a generic integration guide without specific Miles instructions, so the agent had to figure out the integration points by itself.
Agent~B has \sys masked. It may modify Miles and SGLang, and must continue testing concrete hypotheses on both workloads until no remaining idea is expected to improve either AGST or EWTT while preserving correctness.
The agents are asked to measure the EWTT performance for their final implementation.

\begin{table}[H]
  \centering
  \setlength{\tabcolsep}{3.5pt}
  \renewcommand{\arraystretch}{1.08}
  \resizebox{0.4\textwidth}{!}{
  \begin{tabular}{@{}llrrrr@{}}
    \hline
    \textbf{Workload} & \textbf{Variant} & \textbf{EWTT (s)} & \textbf{EWTT speedup} \\
    \hline
    30B-A3B & Baseline & 3.57  & $1.00\times$ \\
    30B-A3B & Agent A & 0.18 & $19.7\times$ \\
    30B-A3B & Agent B & 2.51 & $1.42\times$ \\
    \hline
    32B & Baseline & 1.69 & $1.00\times$ \\
    32B & Agent A  & 0.19 & $8.84\times$ \\
    32B & Agent B & 1.71 & $0.99\times$ \\
    \hline
  \end{tabular}}
  \captionsetup{labelfont=bf}
  \caption{\textbf{Agent-produced results on both frozen four-node workloads.}
  Speedup is over the corresponding broadcast baseline. All data points pass the correctness checks.}
  \label{tab:optimization-easiness}
\end{table}

Table~\ref{tab:optimization-easiness} shows the performance results. Agent~A succeeded on its first attempt to integrate \sys and produced near-optimal EWTTs.
Agent~B profiled and optimized the native broadcast path.
It landed optimizations including shortening lock polling, moving per-bucket rollout confirmations out of the broadcast critical sections, and replacing per-expert-tensor collectives on 30B-A3B with one combined exchange per bucket.
These control plane changes slightly improved EWTT, but failed to address the core redundant transfer issue --- Agent~B identified this redundancy but chose not to optimize it, concluding that the change would require too much code change and be too risky to debug.
It also tried the P2P mechanism that Miles already provided, but rejected that path due to the additional D2H overhead it observed during profiling.
As a result, Agent~B stopped ignore the instruction and stopped before exhausting the optimization space.

Excluding experiment waits, Agent~A spent 24.5 minutes on integration, while Agent~B spent 131 minutes on analysis and optimization; including experiment waits, their total wall times were 73 and 329 minutes, respectively.

\noindent\textbf{Comparison across synchronization modes.} 
\sys uses the sender- and receiver- staging switches to select synchronization modes.
We use the 30B workload from \S~\ref{sec:real-workload} and measure AGST under different stagin combinations.
Fully synchronous transfer achieves the lowest AGST at 137.5\,ms. Enabling staging only on the rollout side raises AGST to 403.4\,ms, enabling it only on the trainer side raises AGST to 441.6\,ms, and enabling staging on both sides for fully asynchronous transfer raises AGST to 609.1\,ms. 
Although staging can overlap inter-node communication with computation, the additional PCIe transfers between host and GPU memory outweigh this benefit on our high-bandwidth testbed. 
Nonetheless, overlapping inter-node traffic with computation can still be beneficial in scenarios where the trainer and rollouts reside in separate clusters, or where a high-speed inter-node network is unavailable.

\subsection{Integration with internal framework}

We integrated \sys with an internal RL framework, which uses a custom weight transfer library.
After each training step, the \intsys first offloads weights from the GPU to a TorchStore volume in CPU memory.
Then each rollout node then fetches from these volumes and aggregates the full model in its own CPU memory before each rollout worker loads the weights it needs.

Although the \intsys uses completely different backend stacks and RDMA library from Miles, a coding agent was able to identify the correct integration points from \sys documentation and replace the original inefficient weight transfer mechanism with \sys's more efficient implementation.
Manual effort was minimal throughout the integration process.

We compare EWTTs between \sys and the internal framework's native path across different numbers of rollout replicas.
Both use the same RDMA backend for cross-node communication.
\sys improves EWTT by $6.29\times$, $7.96\times$, and $11.00\times$ with one, two, and four rollout replicas, respectively.
\sys eliminates the unnecessary communication between hosts and devices and achieves steadily better EWTTs across all scales.
It also scales better than the \intsys due to the removal of the single-source bottleneck.
\sys will have a larger speedup in deployments with larger rollout EP degrees, because the \intsys sends the full model weights to each node, potentially causing redundant transfer.

\subsection{Performance Breakdown}

\begin{figure*}[t]
  \centering
  \captionsetup{skip=3pt}

  \begin{minipage}[t]{0.49\textwidth}
    \begin{minipage}[t][5.05cm][t]{\linewidth}
      \centering
      \includegraphics[height=4.8cm]{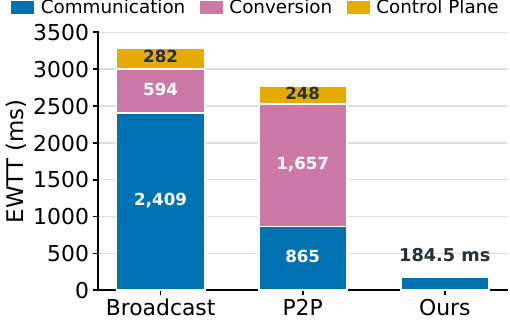}
      \vfill
    \end{minipage}
    \captionof{figure}{\textbf{EWTT breakdown.}
      Numbers are from one representative weight transfer.
      ``Communication'' measures the union of intervals during which RDMA
      fabrics are active.
      ``Conversion'' includes exposed format conversion on both sides and
      the device-to-host delay for Miles P2P.
      ``Control Plane'' includes CPU computation and RPC calls not hidden
      by communication.
      \sys hides all other operations under communication.}
    \label{fig:breakdown}
  \end{minipage}\hfill
  \begin{minipage}[t]{0.49\textwidth}
    \begin{minipage}[t][5.05cm][t]{\linewidth}
      \centering
      \includegraphics[height=4.8cm]{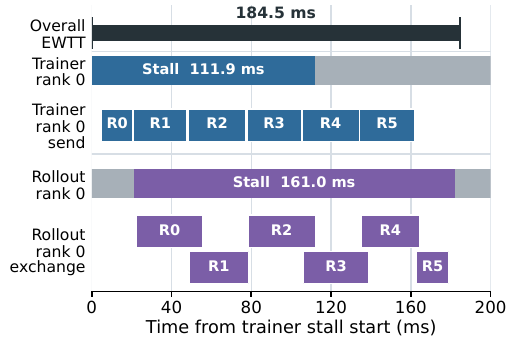}
      \vfill
    \end{minipage}
    \captionof{figure}{\textbf{\sys weight-transfer timeline.}
      R0--R5 mark the six transfer rounds.
      All non-RDMA operations are overlapped.}
    \label{fig:wb-timeline}
  \end{minipage}
\end{figure*}

To better understand our source of gains and the difference to ideal EWTT, we carefully examine the EWTT breakdowns of Miles Broadcast, Miles P2P and \sys under the 30B configuration in \S~\ref{sec:real-workload}.

As shown in Figure~\ref{fig:breakdown}, both Miles Broadcast and P2P expose significant non-communication overhead stemming from control plane operations and format conversion. These overheads even surpass the ideal EWTT due to tensor-by-tensor processing.

Miles P2P largely reduces communication overhead in broadcast by reducing redundancy (see Appendix~\ref{app:breakdown}). 
However, the gains are largely overshadowed by the additional device-to-host transfer overhead in conversion, while the single-source bottleneck and pre-gathering make the communication time still far from the ideal EWTT.

Figure~\ref{fig:wb-timeline} further shows the timelines for selected ranks in one weight transfer of \sys.
\sys uses a depth-2 buffer system to fully overlap the intra-node operations -- such as packing, assembling and consumption -- with the inter-node RDMA communication.
\sys also removes the single-source bottleneck of P2P through deduplication and exchangs among the rollout workers, allowing  EWTT to approaches ideal.
The two-buffer design also lets the trainer worker exit after packing data for R5, shortly after R3 frees the buffer, and lets rollout workers process adjacent rounds in parallel to further saturate RDMA bandwidth.

\section{Discussion}
\label{sec:discussion}

\noindent\textbf{Parameter precision mismatch.} 
\sys assumes that trainer and rollout parameter precisions match.
In practice, some RL recipes use lower-precision parameters for rollouts to speed up inference. 
Quantizing trainer tensors before weight transfer is necessary in such cases.
Quantization algorithms require gathering the full tensor beforehand, which conflicts with the principle that the total trainer egress volume should exactly equal the model size. 
Nevertheless, we believe \sys can be extended to support pre-transfer quantization at the cost of additional trainer-side traffic. 
The correspondence extraction algorithm remains correct. 
In each transfer round, the trainers would gather the sharded tensors to be quantized before issuing the RDMA write. 
We leave this extension to future work.

\noindent\textbf{Elasticity and fault tolerance.}
\sys currently uses trainer-initiated weight transfers with a precomputed plan over a fixed set of workers.
TensorHub~\citep{tensorhub}, RLBoost~\citep{rlboost}, and Laminar~\citep{laminar} explore complementary mechanisms for dynamic resource availability and failure recovery.
TensorHub supports versioned weight retrieval and retries from alternative replicas; RLBoost handles preemptible rollout resources through pull-based provisioning and response migration; Laminar isolates failures through relay workers and reconstructs its dissemination pipeline when relays fail.
These mechanisms are compatible with \sys's architecture.
The transfer planner could be changed to generating plans using a similar relay pattern across rollout replicas as TensorHub, and allow each of them to select its source dynamically.
The transfer planner could generate a plan using a TensorHub-like relaying pattern across rollout replicas and allow each replica to select its source dynamically.
The worker coordinator should also be enhanced to support elastic rollout engine instances.
This extension would support elasticity and fault recovery while preserving efficiency across diverse model layout configurations.

\section{Related Work}
\label{sec:related}

\noindent\textbf{RL frameworks for LLMs.}
TRL~\citep{trl}, DeepSpeed-Chat~\citep{deepspeed}, OpenRLHF~\citep{openrlhf}, and NeMo-Aligner~\citep{nemoaligner} provide scalable infrastructure for model alignment, while RLHFuse~\citep{rlhfuse} improves resource utilization through stage fusion.
VeRL~\citep{verl} and Miles~\citep{miles} integrate independently optimized training and inference engines.
AReaL~\citep{areal}, Relax~\citep{relax}, and AsyncFlow~\citep{asyncflow} explore asynchronous execution; Laminar~\citep{laminar} further decouples training and rollouts through CPU-resident relays for asynchronous weight dissemination.
SkyRL~\citep{skyrl} and prime-rl~\citep{primerl} support modern RL workloads, while StreamRL~\citep{streamrl} and RLBoost~\citep{rlboost} target heterogeneous and preemptible resources, respectively.
RollPacker~\citep{rollpacker}, SpecActor~\citep{specactor}, and StaleFlow~\citep{staleflow} address long-tail rollouts, speculative execution, and staleness-aware coordination.
\sys complements these systems with a weight-transfer layer supporting diverse layouts and synchronization requirements.

\noindent\textbf{Weight-transfer optimizations.}
The growing weight-transfer bottleneck has attracted increasing attention.
Miles-P2P~\citep{milesp2p} prepares destination-format weights using source-side inference-engine replicas, while prime-rl P2P~\citep{primerlwt} traces weight-loading operations to derive weight correspondence.
AWEX~\citep{awex} provides explicit cross-engine weight adaptation.
TensorHub~\citep{tensorhub} exposes versioned weights through reference-oriented storage and uses rollout replicas to serve subsequent requests.
These systems typically show inefficiencies under certain model layout patterns and are designed for only one synchronization mode.
\sys combines automatic layout correspondence extraction, global transfer planning, and configurable synchronization to support diverse RL configurations within one library.

\noindent\textbf{P2P RDMA libraries.}
Mooncake Transfer Engine~\citep{mooncake}, Monarch-RDMA~\citep{monarch}, NIXL~\citep{nixl}, and UCCL~\citep{uccl} provide communication primitives and resource management for efficient transfers between distributed workers.
Fabric-lib~\citep{fabriclib} additionally demonstrates pipelined parameter preparation and one-sided RDMA writes for asynchronous weight trasnfers.
Ray Direct Transport (RDT)~\citep{raydirect} integrates RDMA-backed transfers into Ray and reduces setup overhead through reusable memory registrations and transfer metadata.
These libraries provide the underlying data movement, while heterogeneous layout correspondence, replica-aware routing, and weight-update coordination require additional application-level logic.
\sys currently uses Mooncake as its RDMA backend, with Monarch-RDMA as an alternative, and can accommodate other backends supporting memory registration and direct RDMA writes.
\section{Conclusion}
\label{sec:conclusion}
\sys is a flexible and efficient weight-transfer library for RL post-training. 
It consists of automatic correspondence extractor, the global weight transfer planner, the routing plan executor and the worker coordinator.
These techniques enable \sys to achieves $5$--$42\times$ speedup on weight transfer over existing weight transfer mechanisms across diverse RL configurations.
An coding agent integrated \sys into Miles within 25 minutes without additional manual guidance, showing the simplicity of its APIs.

\section*{Acknowledgements}

We thank Jinghan Sun, Alicia Golden, and Christian Puhrsch for providing general insights on distributed systems for machine learning systems, and Zhenting Qi for insights from an ML researcher's perspective. We are grateful to Tristan Rice, Kapil Sharma and Meet Vadakkanchery for discussion on WeightBridge’s relevance and adoption potential to PyTorch. We also thank Gal Rotem, Danning Xie, and Min Si on discussions over internal post-training systems and infrastructure design, and Neel Raja for setting up access to internal resources.

\clearpage
\bibliographystyle{assets/plainnat}
\bibliography{paper}

\clearpage
\beginappendix
\section{Weight Transfer Overhead is Growing}
\label{app:trend}
\begin{figure}[h]
    \centering
    \includegraphics[width=0.7\linewidth]{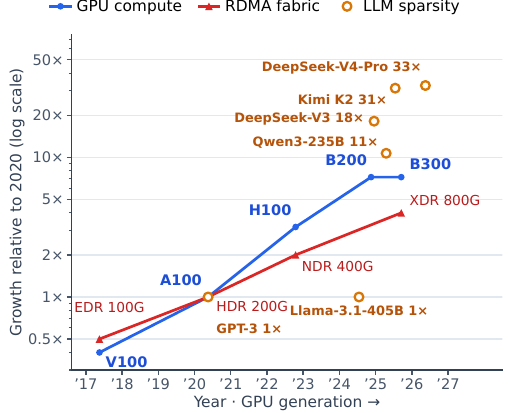}
    \captionsetup{labelfont=bf}
    \caption{\textbf{Scaling Trends of Hardware and Models} --- Byte-normalized dense FLOPS is constant across precisions for a given GPU model. The x-axis shows release dates of actual hardware and models, not specifications.}
    \label{fig:trend}
\end{figure}

For an RL job, the relative overhead of weight transfer equals the communication latency for weight transfer $T_{\text{comm}}$ divided by the total computation time per iteration $T_{\text{comp}}$.
$T_{\text{comm}}$ is proportional to the total model size $M_T$ divided by the communication bandwidth between workers $B$, while $T_{\text{comp}}$ is proportional to the total size of activated parameters per token $M_A$ divided by the GPU's per-byte FLOPS $C$.
Putting these together, $T_{\text{comm}}/T_{\text{comp}}$ is proportional to $\frac{M_T/B}{M_A/C} = \frac{M_T}{M_A}\cdot\frac{C}{B}$. Two important trends are driving this quantity upward, both shown in Figure~\ref{fig:trend}.

\noindent\textbf{Model Sparsity Grows Exponentially.} 
Conventional dense models have encountered bottlenecks in scaling up parameter counts due to prohibitive computation and communication costs. 
As a result, the community has increasingly adopted Mixture-of-Experts (MoE) architectures that activate only a small fraction of the model's parameters per token in a forward pass, significantly reducing computation while preserving the overall model capability.
As larger scale and lower cost are pursued simultaneously, the activated fraction will only continue to shrink. Empirically, $M_T/M_A$ has grown at roughly $92\%$ annually in recent years.

\noindent\textbf{Computing Power Grows Faster than Network Bandwidth.} 
Both GPU computational throughput and interconnect network bandwidth have been growing exponentially in recent years.
However, the growth rate of computational power ($\sim42\%$/yr) outpaces that of RDMA bandwidth ($\sim24\%$/yr).
This yields approximately $14.5\%$ annual growth in the $C/B$ term (see Appendix~\ref{app:trend}).
As cross-datacenter RL becomes more prevalent, driven by the increasing scale of model and RL workloads, this hardware-level disparity will only intensify.

Combining these two trends above, we project that the communication-computation ratio in RL workloads will grow roughly $2.2\times$ annually, introducing, a real urgency for optimized weight transfer mechanisms.

As shown in Figure \ref{fig:trend}, the annual growth rates for LLM sparsity ($M_T/M_A$), byte-normalized FLOPS ($C$) and RDMA fabric bandwidth ($B$) are $\sim92\%$, $\sim 42\%$ and $\sim 24\%$, respectively, giving us a combined $\sim 2.2\times$ annual growth in relative weight transfer overhead.

\section{Details in the source map extraction algorithm}
\label{app:loadspec}

\begin{algorithm}[t]
\small
\caption{Automatic Source Map Extraction}
\label{alg:extract}
\begin{algorithmic}[1]
\Require checkpoint weights $\mathcal{W}_{\mathrm{ckpt}}$, in-GPU parameters
         $\mathcal{W}_{\mathrm{gpu}}$, black-box \texttt{load\_weights},
         chunk width $b\in\{8,16\}$
\Ensure  Loading Specification $\mathcal{L}$

\State $K \gets \lceil 64/b \rceil$
      \Comment{chunks per \texttt{int64} identifier}
\State $\mathcal{H}_{Z} \gets \emptyset$,\;
       $\mathcal{H}_{\Gamma} \gets \emptyset$
      \Comment{host: $\mathrm{diff2}$ records, carry records}

\For{$k \gets 0$ \textbf{to} $K-1$}
  \State $\widehat{C} \gets \textsc{MakeIdChunk}(\mathcal{W}_{\mathrm{ckpt}}, k, b)$
        \Comment{$\lfloor p / 2^{bk} \rfloor \bmod 2^{b}$}
  \State $\texttt{load\_weights}(\widehat{C}, \mathcal{W}_{\mathrm{gpu}})$
        \Comment{opaque}
  \ForAll{tensors $A \in \mathcal{W}_{\mathrm{gpu}}$ \textbf{in parallel}}
    \State $D \gets \mathrm{diff2}(A)$
          \Comment{\textsc{gpu}, Eq.~\eqref{eq:diff2}}
    \If{$k > 0$}
      \State $\Gamma \gets \textsc{ExpandCarry}\bigl(\mathcal{H}_{\Gamma}[k{-}1]\bigr)$
            \Comment{host $\rightarrow$ \textsc{gpu}}
      \State $D \gets D + \Gamma$
            \Comment{cancels wrap artifacts}
    \EndIf
    \State $\mathcal{H}_{Z}[k] \gets \textsc{NonZeros}(D)$
          \Comment{\textsc{gpu} $\rightarrow$ host, $O(N_c)$}
    \State $\mathcal{H}_{\Gamma}[k] \gets \textsc{StridedDiff1}\bigl(\textsc{EmitCarry}(D)\bigr)$
          \Comment{\textsc{gpu} $\rightarrow$ host, $O(N_c)$}
  \EndFor
\EndFor

\State $\mathcal{Z} \gets \textsc{Recombine}(\mathcal{H}_{Z})$
      \Comment{$\textstyle\sum_k 2^{bk}\,\mathcal{H}_{Z}[k]$}
\State $\mathcal{L} \gets \emptyset$
\ForAll{tensors $A \in \mathcal{W}_{\mathrm{gpu}}$}
  \State $\mathcal{L} \gets \mathcal{L} \cup
          \textsc{ScanSliceMappings}(\mathcal{Z}_A)$
        \Comment{host; index tensor never materialized}
\EndFor
\State \Return $\mathcal{L}$
\end{algorithmic}
\end{algorithm}

\textbf{Difference encoding.} Let $A$ denote the identifier field of an in-GPU tensor, where $A_{i,j}$ identifies the position in \texttt{w\_ckpt} that supplies the element at row $i$ and column $j$ of $A$. W.l.o.g. we assume that $A$ is two-dimensional.
We first define the two-dimensional first-order difference
\begin{equation}
\label{eq:diff1}
\mathrm{diff1}(A)_{i,j} \;=\; A_{i,j} - A_{i-1,j} - A_{i,j-1} + A_{i-1,j-1},
\end{equation}
with the convention $A_{i,j}\triangleq 0$ whenever $i<0$ or $j<0$, so that $\mathrm{diff1}$ is invertible by a two-dimensional prefix sum.
The quantity we actually store is its self-composition,
\begin{equation}
\label{eq:diff2}
\mathrm{diff2}(A) \;=\; \mathrm{diff1}\bigl(\mathrm{diff1}(A)\bigr),
\end{equation}
which expands into a $3\times3$ convolution kernel given by the outer product of $[\,1,-2,1\,]$ with itself, and is also lossless.

The \texttt{diff2} operation leaves us with a very sparse tensor.
Each of the $N_c$ slice-copy operations induces an identifier field that is affine---with or without transposition---over the rectangular region it fills. A single application of $\mathrm{diff1}$ leaves non-zeros only along the boundary of each rectangle, and the second application annihilates those boundary ramps as well, confining the surviving non-zeros to a constant-size neighborhood of each rectangle's corners.

As a result, the total number of non-zero entries across the whole model is reduced to $O(N_c)$ rather than $O(M)$ through \texttt{diff2}, so we store and transmit to the CPU only these non-zero entries.

However, chunking and difference encoding interact badly when composed: within a lower-order chunk, the \texttt{diff2} computation on truncated indices will very likely overflow, and these carried values must be applied to the next chunk for correctness.

We observe that these carry positions are themselves regularly spaced, being induced by the fixed truncated width of indices.
We therefore compute an additional strided difference over the \texttt{diff2} tensors, which cancels the periodic carry artifacts and encodes the carried bits explicitly, restoring the $O(N_c)$ bound on retained entries.

\textbf{The full algorithm.} Algorithm \ref{alg:extract} summarizes the source map generation algorithm described in \S~\ref{sec:correspondence}.

\section{Detailed breakdown for Miles broadcast and P2P}
\label{app:breakdown}

\begin{figure*}[t]
  \centering
  \includegraphics[width=0.9\textwidth]{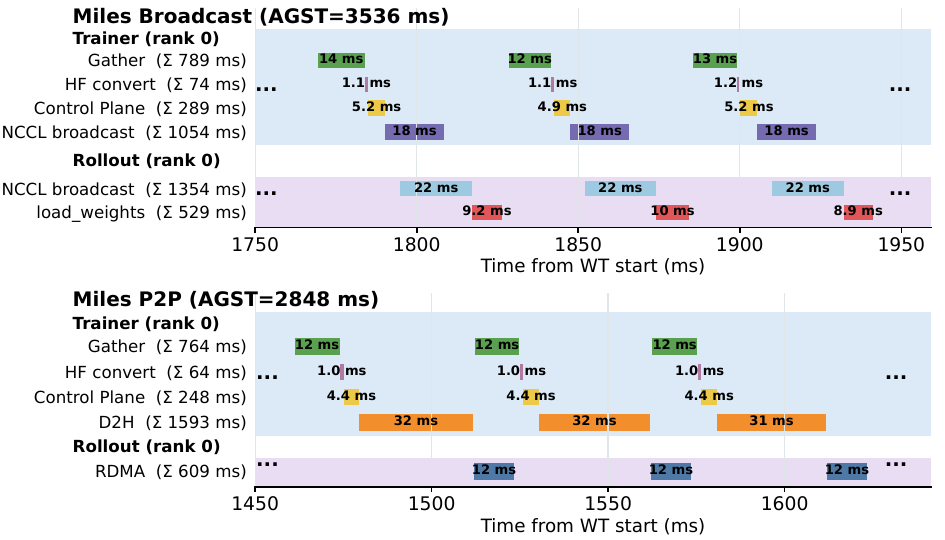}
  \captionsetup{labelfont=bf}
  \caption{\textbf{Zoomed in performance breakdown of baselines.}
  Miles broadcast and P2P execute weight transfer in batches of tensors.
  Bars show per-operation time spans for three selected consecutive batches under the 4-node (30B) configuration in Figure~\ref{fig:real-workload}.
  We select one worker from each of the trainer engine and one from each inference engines.
  $\Sigma$ covers all 58 batches.
  Both NCCL-broadcast rows bracket collective submission and handle waits; the rollout interval additionally includes receive-tensor allocation.
  CPU Prep mainly covers control plane work including per-tensor processing times and Ray RPC call latencies.
  Miles P2P's RDMA interval is source-timed but writes directly into the labeled rollout target.
  }
  \label{fig:miles-timeline}
\end{figure*}

To provide a deeper insight into the performance bottlenecks of the baselines, we show detailed timeline breakdowns of both Miles broadcast and Miles P2P in Figure~\ref{fig:miles-timeline}.

\section{Varying RDMA Buffer Size}

\begin{figure}[t]
  \centering
  \includegraphics[width=0.7\linewidth]{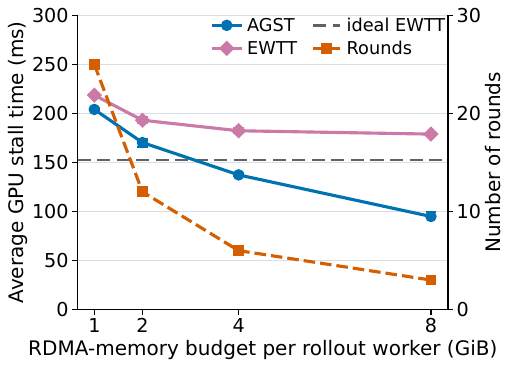}
  \captionsetup{labelfont=bf}
  \caption{\textbf{Varying RDMA buffer size.}
  Results of the 30B model configuration in \S~\ref{sec:real-workload} with different registered RDMA-memory budget per rollout worker. The default configuration is 4\,GiB.
  }
  \label{fig:sensitivity-buffer}
\end{figure}

\sys takes a pre-configured RDMA buffer size and computes the number of transfer rounds needed to fit within that budget. 
We vary the RDMA memory budget for the 30B setting in \S~\ref{sec:real-workload} and measure the AGST. 
Figure~\ref{fig:sensitivity-buffer} shows the results.
As the memory budget increases, EWTT gradually converges to a value slightly above ideal. 
This is because the number of rounds decreases, which in turn reduces control-plane delays and pipeline bubbles. 
AGST decreases even faster, since the head and tail of the 3-stage pipeline account for a larger fraction of the EWTT as the number of rounds drops.

\end{document}